\documentclass[11pt]{article}
\usepackage{hyperref}
\usepackage{graphicx,amssymb,amsmath,amsfonts,cite}
\usepackage[margin=0.75in]{geometry}
\usepackage{color}
\usepackage[affil-it]{authblk}

\newcommand{\cE}{{\cal E}}

\newcommand{\cK}{{\cal K}}  
  
\newcommand{\cO}{{\cal O}}

\newcommand{\cU}{{\cal U}}

  \newcommand{\bbP}{{\mathbb P}}

\newcommand{\tF}{\tilde{F}}

\newcommand{\tilh}{{\tilde{h}}}

\newcommand{\be}{\begin{equation}} \newcommand{\ee}{\end{equation}}
\newcommand{\bea}{\begin{eqnarray}} \newcommand{\eea}{\end{eqnarray}}
\newcommand{\beann}{\begin{eqnarray*}}  \newcommand{\eeann}{\end{eqnarray*}}
\newcommand{\bfig}{\begin{figure}} \newcommand{\efig}{\end{figure}}
\newcommand{\ba}{\begin{array}} \newcommand{\ea}{\end{array}}
\newcommand{\bcen}{\begin{center}} \newcommand{\ecen}{\end{center}}
\newcommand{\btab}{\begin{tabular}} \newcommand{\etab}{\end{tabular}}

\def\d{\partial}

\title{Unparticles as the Holographic Dual of Gapped AdS Gravity}
\author{Sophia K. Domokos\thanks{skd5@nyu.edu} ~ and~ Gregory Gabadadze\thanks{gg32@nyu.edu}}
\affil{Center for Cosmology and Particle Physics\\ Department of Physics, New York University\\  New York, NY 10003, USA}
\date{\vspace{-7ex}}

\begin{document}

\maketitle

\abstract{Naively applying holographic duality to
gapped gravity on Anti de Sitter (AdS) space seems to suggest that the stress tensor 
of the field theory dual  cannot be conserved. On the other hand, by symmetry arguments, it seems that the dual 
should not violate Poincare symmetry. To clarify this apparent contradiction,  
we study a holographic dual of  massive gravity  where both
the physical background metric and the fiducial metric are AdS. Using the anomalous scaling of the energy momentum tensor as our guide, we conclude that the dual theory is nonlocal. We find that the dual  is  
similar to  conformal invariant ``unparticle" theories. We show that such theories  can be viewed as dimensional reductions of flat-space field theories 
with  inhomogeneous scaling properties. }

\section{Introduction}

While the proposal  that gravity  is holographic  in its nature is profound and mysterious  \cite{'tHooft:1993gx, Susskind:1994vu}, it does have a concrete realization:
 the AdS/CFT correspondence \cite{Maldacena:1997re,Gubser:2002tv,Witten:1998qj}.
How  would this correspondence manifest itself  if conventional
General Relativity (GR) were continuously deformed into  a different theory?
Massive gravity on AdS space represents a playground in which we can explore this question.  

The nonlinear theory of a massive graviton -- referred to here as mGR -- 
was  constructed in \cite{deRham:2010ik,deRham:2010kj}. The theory guarantees that in $(d+1)$-dimensional spacetimes, the graviton mode has  $(d+2)(d-1)/2$  
physical polarizations (instead of GR's $(d+1)(d-2)/2$) on any sensible background  \cite {deRham:2010ik,deRham:2010kj,Hassan:2011hr,Hassan:2011tf, Hassan:2011ea,Deffayet:2012nr,Deffayet:2012zc,Bernard:2015mkk,Mirbabayi:2011aa,Hinterbichler:2012cn} (e.g. AdS).
Furthermore, the $m\to 0$  limit of a linearized massive gravity  on AdS space is 
continuous \cite{Porrati:2000cp}. Thus, assuming that mGR has  a stable asymptotically AdS 
solution for the metric --- as we show is the case --- it is interesting to wonder  what kind of 
holographic correspondence  could exist between mGR on AdS$_{d+1}$ spacetime and some 
$d$-dimensional field theory.

The solution to the mGR equations of motion generically involves 
background values for the St{\"u}ckelberg fields, which carry the additional degrees of freedom that appear due to the graviton mass.
Such solutions in general  map to field theory vacua with non-trivial vacuum expectation values (vevs) or even explicit sources.  
Since such vevs would break Poincare symmetry, the field theory's energy-momentum tensor (EMT) should not be conserved. Such a construction has already appeared in the AdS/CMT literature. \cite{Vegh:2013sk, Blake:2013owa} for example, break translational invariance in the dual theory by turning on an inhomogeneous reference metric, which corresponds to adding a lattice of ions in the otherwise homogeneous field theory. 

This statement is not surprising if we follow standard AdS/CFT lore. Gauged symmetries on AdS correspond to global symmetries in field theory, and AdS gauge fields  are dual to conserved currents. Giving the gauge field a mass in AdS breaks the corresponding global symmetry in the field theory, generating a nozero divergence for the field theory current. Since the massless graviton is dual to the EMT, turning on a mass for the graviton should ruin EMT conservation. While this is indeed the case for  \cite{Vegh:2013sk, Blake:2013owa, Nomura:2014uja}  and others in theories that \textit{explicitly} break some of the conformal invariance, there exists a very simple example for which the interpretation is not so clear. 

The covariant formulation of mGR necessitates a ``reference metric'' $f_{ab}$, which defines the geometry of the target space parametrized by St{\"u}ckelberg fields $\Phi^a$. It is the choice of $f_{ab}$ which defines -- in part -- which of a large set of possible mGR theories we are studying. Say we choose the reference metric $f_{ab}$ to be AdS, and work in a vacuum where $\Phi^a=\delta^a_M x^M$. (Uppercase indices denote spacetime indices, lowercase target space indices.) In that case, a pure AdS metric $g_{MN}$ with no St{\"u}ckelbergs exactly solves the nonlinear mGR equations of motion. 
 AdS$_{d+1}$ with massive graviton represents a valid vacuum state. Since the solution maintains the symmetries of AdS, the dual field theory should still enjoy unbroken conformal invariance. Yet the graviton is massive: its dual operator now displays unusual scaling properties, which naively implies a violation of Poincare symmetry.

Addressing this puzzle is the focus of the present note. First, we describe the holographic interpretation of mGR in general terms. We then focus on the operator content of its dual conformal field theory (CFT), using the the $m\rightarrow 0$ limit as a guide.  We then analyze two-point functions in the $m\ne 0$ theory, concluding eventually that the dual  theory must be nonlocal. We furnish an example of such a theory:  the ``unparticle stuff''  of \cite{Georgi:2007ek}. We illustrate  how such nonlocal conformal theories could arise as dimensional reductions of flat space theories. Finally, we conclude and suggest some directions for future work.

\section{Massive gravity and duality}

Let us begin by outlining the general structure of mGR theories, and how to interpret it in the context of holography.

In $(d+1)$ dimensions, mGR possesses $(d+2)(d-1)/2$ physical degrees of freedom (DOF). We can package all of these DOF into fluctuations of the massive metric $g_{MN}$, but it is often more convenient to introduce a gauge redundancy that restores diffeomorphism invariance. We can thus define mGR in terms of the bulk metric $g_{MN}$ \textit{and} $(d+1)$ St{\"u}ckelberg fields $\Phi^a$. The reference metric $f_{ab}$ is a function of these $\Phi^a$, which transform as scalars under diffeomorphisms. One can then construct diffeomorphism invariant terms using the object
$f_{ab}\d_M \Phi^a\d_N \Phi^bg_{MN}$.  Hence mGR possesses an additional internal symmetry due to the reference metric $f_{ab}$, and  is \textit{not} simply a deformation of massless GR. As we see below, even the zero-mass limit of mGR is not identical to standard GR. We can, instead, think of massive gravity as a non-linear sigma model (NLSM) in the presence of bulk gravity $g_{MN}$. The reference metric $f_{ab}$ defines the metric on the target space, while the  St{\"u}ckelbergs $\Phi^a$ denote target space coordinates. The symmetries of $f_{ab}$ then correspond to internal symmetries: when $f_{ab}$ is AdS$_{d+1}$, for instance, the target space of the NLSM features a non-linearly realized $SO(2,d)$  symmetry that transforms the $\Phi^a$ among themselves.  

Consider background values for the St{\"u}ckelbergs,  $\Phi^a\delta^M_a=x^M$ which break the spacetime diffeomorphism invariance and the reparametrization invariance of $f_{ab}$ to the diagonal subgroup. We now have 
$f_{ab}\d_M\Phi^a\d_N\Phi^bg_{MN}\equiv f_{MN}g^{MN}$, and solve Einstein's equations in this background. A generic solution could have a non-trivial $g_{MN}$ \textit{and} non-vanishing fluctuations of the St{\"u}ckelberg fields $\delta\Phi^a$. 
However, it turns out that taking $g_{MN}$
AdS$_{d+1}$, with $\delta \Phi^a=0$  also solves the equations of motion. It is straightforward to see that this is the case. The full non-linear action takes the form \cite {deRham:2010kj}
\begin{align}\label{eq:MGRaction}
S=M_P^{d-1}\int d^{d+1}x\sqrt{-g}\left[ R(g) - 2\Lambda_{cc}+2m^2 \cU(\cK^M_N)\right] + S_{GHY}+S_{ct}
\end{align}
where 
\begin{align}
\cK^A_{\phantom{A}B} = \delta^A_B - \sqrt{g^{-1}f}^{~A}_{\phantom{A}B},~~~~f_{MN} = 
f_{ab}\d_M\Phi^a\d_N\Phi^b\,,
\end{align}
for bulk metric $g_{MN}$ and reference metric $f_{MN}$. $\Lambda_{cc}$ is the cosmological constant. The potential $\cU(\cK)$ is a  sum of all the determinants of the matrix $\cK$,  from  $det_2(\cK)$, to $det_{d+1}(\cK)$ with arbitrary coefficients. 
$\sqrt{g} det_1(\cK)$ can be expressed -- up to a total derivative -- as a linear combination of the cosmological constant and the 
other determinants.  

Meanwhile, $S_{GHY}$ is the usual Gibbons-Hawking-York boundary term of massless GR. There are no additional boundary terms needed for a consistent variational principle: the invariants in the potential $\cU(\cK^M_N)$ contain at most one derivative per field, 
so their variations can be eliminated with boundary conditions alone. 
$S_{ct}$ refers to a set of boundary counterterms that are needed to regularize the near-boundary divergence of AdS/CFT correlation functions.

For \textit{any} reference metric $f_{MN}$ that solves the massless Einstein equation,  taking $g_{MN}=f_{MN}$ reduces the massive gravity equations  of motion to the massless Einstein equation, which $f_{MN}$ satisfies by definition.  We can then consider weakly coupled fluctuations in this background as long as the energy scale of fluctuations is below some scale $\Lambda_{strong}$ on the gravity side (we return to this point briefly in our concluding remarks). We will take both $f_{MN}$ and $g_{MN}$ to be Anti de Sitter. This solution not only solves the equations of motion, but is stable under small fluctuations, as it reduces in the linearized 
approximation  to the Fierz-Pauli quadratic Lagrangian  on an AdS  background (see \cite {Porrati:2000cp}, 
and references therein).

We can now ask: what does this story looks like in holography? 

 From the perspective of the dynamical metric $g_{MN}$, there is simply no holographic description until we choose the AdS$_{d+1}$ vacuum for $g_{MN}$. Once we fix this background, we have a field theory $CFT_g$, with a spin $2$ current (i.e. the EMT), a spin 1 current, and a scalar operator. The $n$-point functions of these operators are fully determined by the action \eqref{eq:MGRaction}. The action depends on the fact that $f_{ab}$ is AdS, so the NLSM of the $\Phi^a$ fields 
has an internal  conformal symmetry. 
 However, the $\Phi^a$ fields have no kinetic  
terms  if the theory is expanded around  $\langle\Phi^a\rangle =0$; thus, the gravity theory is infinitely strongly 
coupled at the origin in the field space. In other words, holography does not afford us any advantages at this point.
On the other hand, if  we choose a vacuum in which $\Phi^a=\delta^a_M X^M$, which defines a non-trivial vev for some operators in $CFT_g$, and there is an ``emergent'' weakly coupled theory. The conformal symmetry and internal Poincare symmetry are now broken to a diagonal subgroup, as the background solution explicitly mixes the internal and bulk degrees of freedom.

From now on we focus  on the dual field theory of massive gravity with  $\Phi^a=\delta^a_M x^M$ and 
background metric AdS$_{d+1}$, under which we have $f_{ab}\d_M\Phi^a\d_N\Phi^b\equiv f_{MN}=g_{(0)MN}$.

\section{The massless limit: counting degrees of freedom and identifying dual operators}
 
Let us now turn to a more technical study of the operator content in the $d$-dimensional flat space field theory.  We have taken the reference metric  $f_{ab}$ to be AdS$_{d+1}$. As shown above, pure AdS$_{d+1}$ with no St{\"u}ckelbergs exactly solves 
the massive gravity equations of motion, so we also take $g_{MN}$ to be AdS$_{d+1}$:
\begin{align}\label{eq:AdSmetric}
ds^2=g^{(0)}_{MN}dx^Mdx^N = \frac{L^2}{z^2}\left(\eta_{\mu\nu} dx^\mu dx^\nu + dz^2\right)~.
\end{align}
We work in the mostly plus convention. Uppercase Latin indices denote coordinates in AdS$_{d+1}$; lowercase Greek indices denote coordinates in $d$-dimensional Minkowski space. $z\in [0,\infty)$ is the radial AdS coordinate, 
with boundary at $z=0$. $L$ is the AdS radius. 

We now study fluctuations $h_{MN}$ around the background $g_{(0)MN}$. 
The quadratic order action for metric fluctuations is simply the Fierz-Pauli action on AdS$_{d+1}$:
\begin{align}\label{eq:SFPoriginal}
S_{FP}=\frac{M_P^{d-1}}{4}\int d^{d+1}x\sqrt{-g^{(0)}} \left[ -h^{MN}\mathcal{E}_{MNPQ}h^{PQ} -\frac{1}{2}m^2\left( h_{MN}h^{MN}-h^2\right)  \right]~,
\end{align}
where $h\equiv h^M_M$, all indices are raised and lowered with the AdS metric $g^{(0)}_{MN}$ \eqref{eq:AdSmetric}, and $\mathcal{E}$ is the Einstein (Lichnerowicz) operator \cite{deRham:2012kf},
\begin{align}
({\mathcal E}_{AdS}h)_{MN}=&-\frac{1}{2}\left[\nabla^2\left(h_{MN}-g^{(0)}_{MN}h\right)-\left(\nabla_A\nabla_N h^A_M+\nabla_A\nabla_M h^A_N\right)\right.\nonumber
\\&\qquad\qquad+\left.\nabla_M\nabla_N h+g^{(0)}_{MN}\nabla_A\nabla_Bh^{AB} -\frac{2d}{L^2}\left(h_{MN}-\frac{1}{2}g^{(0)}_{MN}h\right) \right]~,
\end{align}
 and $M_P$ is the $(d+1)$-dimensional Planck mass. $h_{MN}$ is a symmetric two-tensor in $(d+1)$ dimensions. By naive counting it has  $\frac{1}{2}(d+2)(d+1)$ independent components, but the linearized
 equations of motion impose $d+2$ constraints, so on shell there are a total of
\begin{align}
n_h=\frac{1}{2}(d+1)(d+2)-(d+1)-1=\frac{1}{2}(d+2)(d-1)
\end{align}
physical  DOF.  The mGR theories described in \cite{deRham:2010ik,deRham:2010kj}  with action \eqref{eq:MGRaction} maintain this number of degrees of freedom in the full nonlinear theory, even on arbitrary backgrounds.   
These degrees of freedom should precisely match the operator content of the  corresponding CFT.

As noted above, it can be helpful to decompose the graviton mode in a way that restores the diffeomorphism invariance of the theory. We thus take
\begin{align}\label{eq:hdecomp}
M_P^{\frac{d-1}{2}}h_{MN}=\bar{h}_{MN}+\alpha L\nabla_{(M} \bar{A}_{N)}+\beta L^2\nabla_M\nabla_N \bar{\pi}~,
\end{align}
where the $\bar{A}_M$ and $\bar{\pi}$ are the St{\"u}ckelberg fluctuations in the vacuum $\Phi^a=\delta^a_M x^M$, and $\alpha$, $\beta$ are parameters to be determined.
Since the decomposition takes the form of a diffeomorphism $h_{MN}\rightarrow h_{MN}+\nabla_{(M}\xi_{N)}$,  the 
Einstein-Hilbert action is unaffected; all new terms quadratic terms come from the $m^2$ term. The linearized action becomes
\begin{align}\label{eq:SFPdecomp}
S_{FP}&=\frac{1}{4}\int d^{d+1}x\sqrt{-g^{(0)}} \left\{  -\bar{h}^{MN}\mathcal{E}_{MNPQ}\bar{h}^{PQ}\right.\nonumber\\ 
&\qquad-\frac{1}{2}m^2\left[\left( \bar{h}_{MN}\bar{h}^{MN}-\bar{h}^2\right)+\frac{\alpha^2 L^2}{4}\bar{F}_{MN}\bar{F}^{MN}+d\alpha^2\bar{A}^M\bar{A}_M+d\beta^2 L^2\nabla_M\bar{\pi}\nabla^M\bar{\pi}\right.\nonumber\\
&\qquad\qquad\qquad+2\alpha L\left(\bar{h}_{MN}\nabla^M\bar{A}^N
- \bar{h}\nabla^M\bar{A}_M\right)+ 2\beta L^2\bar{h}^{MN}\nabla_M\nabla_N\bar{\pi}-\beta\bar{h}\nabla^2\bar{\pi}\nonumber\\
&\qquad\qquad\qquad+\left.\left.\left.\alpha\beta\frac{d}{L^2}\bar{A}^N\nabla_N\bar{\pi}\right)\right] \right\}~,
\end{align}
where $\bar{F}_{MN}=\d_M\bar{A}_N-\d_N\bar{A}_M=\nabla_M\bar{A}_N-\nabla_N\bar{A}_M$. We have used the fact that  the Ricci tensor on $AdS_{d+1}$ is given by $R_{MN}=-\frac{d}{L^2}g_{(0)MN}$. 
A conformal transformation allows us to eliminate cross kinetic terms between $\bar{h}_{MN}$ and $\bar{\pi}$:
\begin{align}
 \bar{h}_{MN}\rightarrow\tilh_{MN}+ \beta \frac{(mL)^2}{d-1} g_{(0)MN}\bar{\pi}.
 \end{align}
 Appropriately choosing $\alpha$, $\beta$ 
yields canonically normalized kinetic terms for $\bar{\pi}$ and $\bar{A}_M$, so the original graviton  in terms of $\tilh$, $\bar{A}$, $\bar{\pi}$  is now
\begin{align}\label{eq:fullh}
h_{MN}&=M_P^{-(d-1)/2}\left[ \tilh_{MN} + \frac{2\sqrt{2}}{m}\nabla_{(M}A_{N)}+\frac{2\sqrt{d-1}}{\sqrt{d}}\frac{1}{m(L^{-1})\sqrt{(mL)^2+d-1}}\nabla_M\nabla_N\bar{\pi} \right.\nonumber\\
&\qquad\qquad\qquad\qquad\qquad\qquad\qquad\qquad\qquad\qquad+\left. \frac{2}{\sqrt{d(d-1)((mL)^2+d-1)}}g_{(0)MN}\bar{\pi}\right]~,
\end{align}
with linearized action
\begin{align}
S_{FP}=&\int d^{d+1}x\sqrt{g_{(0)}}\left\{ \frac{1}{4}\tilh^{MN}(\mathcal{E}\tilh)_{MN} -\frac{m^2}{8}\left( \tilh_{MN}\tilh^{MN}-\tilh^2\right)\right.\nonumber\\
&\qquad\qquad\qquad\qquad-\frac{1}{4}\bar{F}_{MN}\bar{F}^{MN}-\frac{1}{2}\frac{2d}{L^2} \bar{A}_M\bar{A}^M -\frac{1}{2}\d_M\bar{\pi}\d^M\bar{\pi}+ \frac{1}{2}\frac{d+1}{d-1}m^2\bar{\pi}^2\nonumber\\
&\qquad\qquad\qquad\qquad-\frac{m}{\sqrt{2}}\tilh^{MN}\left( \nabla_M \bar{A}_N-g_{(0)MN}\nabla_P\bar{A}^P\right) +\frac{m}{2L} \sqrt{\frac{d(d-1+(mL)^2)}{d-1}}\tilh\bar{\pi}\nonumber\\
&\qquad\qquad\qquad\qquad+\left.\frac{1}{L}\sqrt{\frac{d(d-1)}{2(d-1+(mL)^2)}}\bar{\pi}\nabla_P\bar{A}^P\right\}~.
\end{align}
Note that the action contains three different scales: $M_P$, $m$, and $L^{-1}$. For generic values of these scales, the spin 2, spin 1, and spin 0 parts mix nontrivially. 
In  the massless limit on AdS (when $m\rightarrow 0$, $L^{-1}$ small but finite), however, the spin 2 piece decouples completely from the spin 1 and spin 0 fields, and 
the spin 1 and spin 0 fields then combine into a single massive vector on AdS
with mass $m_V^2 = 2dL^{-2}$. There is no limit in which this vector is massless, so this set of theories always includes the same number of degrees of freedom, regardless of the value of $m$.

Let us now turn to the CFT. Unlike the case of the standard AdS/CFT correspondence, where string theory helped identify both the large $N_c$ CFT and the corresponding supergravity theory, here we can, at best, characterize our CFT in terms of
its symmetry structure and correlation functions, viewing it perhaps as an effective description of a more complicated UV theory. 
Let us say, then, that the operator dual to $h_{MN}$ is some tensor $\tau_{\mu\nu}$. A symmetric tensor in $d$ dimensions would have
$\frac{d(d+1)}{2} $
components, one more than the number of DOF in  the bulk field $h_{MN}$. $\tau_{\mu\nu}$ should thus suffer one additional constraint. Only two possibilities do not violate Lorentz  invariance:
\begin{align}\label{eq:conditions}
\tau_{\mu\nu}\eta^{\mu\nu}=0\qquad\text{and}\qquad\d^\mu\d^\nu\tau_{\mu\nu}=0~.
\end{align}
In order to see that $\tau_{\mu}^{\mu}=0$ is the only viable option, we consider the tensor and vector parts of $\tau_{\mu\nu}$  dual to $\tilde{h}_{MN}$ and $A_M$, respectively. These modes have scaling dimension $\Delta_T$ for tensors
and $\Delta_J$ for vectors. Using the well-known AdS/CFT relations between mass and scaling dimension, we have
\begin{align}
\text{vector:}&\qquad \Delta_J=\frac{1}{2}\left( d+\sqrt{(d-2)^2+4(m_VL)^2} \right)~,\\
\text{spin 2:}&\qquad \Delta_T=\frac{1}{2}\left( d+\sqrt{d^2+4(m_hL)^2} \right)~.
\end{align}
As $m\rightarrow 0$ limit, $\tilde{h}_{MN}$ becomes massless, so the tensor part of $\tau_{\mu\nu}$ must correspond to a dimension $d$ operator, the EMT $T_{\mu\nu}$. Meanwhile, the  CFT current $J_\mu$  dual to $A_M$ has $\Delta_J=d+1$  when $m=0$. Thus, the appropriate $m\rightarrow 0$ limit consists of a theory with conserved EMT, \textit{and} a non-conserved current $J_\mu=(J_\mu^T, \cO)$ with large ``anomalous'' scaling dimension. We still need to impose one of the conditions \eqref{eq:conditions} in order to get the right number of degrees of freedom. $J_\mu$ unequivocally contains $d$ DOF, so we need to effectively impose the extra restriction on $T_{\mu\nu}$. When $m\rightarrow 0$, $\d^\mu T_{\mu\nu}=0$, so imposing 
$\d^\mu\d^\nu T_{\mu\nu}=0$ does not get rid of any additional components. For consistency with the massless limit, then, $\tau^\mu_\mu=0$ is our only choice.\footnote{Indeed, for a conformally invariant local theory, we should have $T^\mu_\mu=0$.}

How, then, do $T^{Tt}_{\mu\nu}$, $J^T_\mu$ and $\cO$ fit into $\tau_{\mu\nu}$? Since $\tau$ must be traceless, we can only write
\begin{align}\label{eq:taudecomp}
\tau_{\mu\nu}&=T^{Tt}_{\mu\nu}+a\frac{1}{\d^2}\d_{(\mu}J^T_{\nu)}+b\frac{1}{2}\frac{1}{\d^2}\left( d\d_\mu\d_\nu-\eta_{\mu\nu}\d^2\right)\cO~,
\end{align}
with $\d^\mu J^T_\mu=0$. Note
the factors of $1/\d^2$  in this definition, which are essential to making sure that all parts of $\tau$ transform in the same way under dilatations. For the same reason, we cannot use a constant scale to soak up the dimensions
of the $J_\mu$  (with dimension $d+1$) appearing in the decomposition of $\tau_{\mu\nu}$  (with dimension $d$). 

The parameters $a,~b$ are arbitrary so far. We can determine them for small $mL$, by requiring that $T^{Tt}_{\mu\nu}$ decouples from $J_\mu$ as $m\rightarrow 0$.  
Consider the source term in the CFT generating functional for $\tau_{\mu\nu}$:
\begin{align}
Z_{CFT}=\exp\left\{i S_{CFT}+i\int d^dx h_0^{\mu\nu}\tau_{\mu\nu}\right\}~.
\end{align}
$h_{0\mu\nu}$ is the boundary value of the $(d+1)$-dimensional graviton, which we have now decomposed in terms of the massless graviton  $\tilde{h}_{MN}$  with boundary values $\tilde{h}_{0\mu\nu}$, and the St{\"u}ckelbergs  $\bar{A}_M$ and $\bar{\pi}$ which constitute a massive gauge field having boundary values $A_{0\mu}$  and $\bar{\pi}_0$. Using equations \eqref{eq:fullh} and \eqref{eq:taudecomp}, we can write the coupling of the boundary values of the AdS$_{d+1}$ fields to the dual operator $\tau_{\mu\nu}$.
 Rewriting the source term appearing in the partition function (after integration by parts) :
\begin{align}
-i\log Z_{CFT}&=S_{CFT}+\int d^dx h_0^{\mu\nu}\tau_{\mu\nu}\nonumber\\
&=S_{CFT}+\int d^dx M_P^{-(d-1)/2}\left\{ \tilde{h}_0^{\mu\nu}\left(T_{\mu\nu}^{Tt} +a\frac{1}{\d^2}\d_{(\mu}J^T_{\nu)}+\frac{b}{2}\frac{1}{\d^2}\left( d\d_\mu\d_\nu-\eta_{\mu\nu}\d^2\right)\cO
 \right)\right.\nonumber\\
 &\qquad\qquad\qquad\qquad + \left.\frac{a\sqrt{2}}{m}A_{0\mu}J^{T\mu}+\frac{\sqrt{2}(d-1)b}{m}A_{0}^\mu\d_\mu\cO+\frac{(d-1)^{3/2}b}{mL^{-1}\sqrt{d(d-1+(mL)^2})}\bar{\pi}_0\d^2\cO\right\}~.
\end{align}
For small $mL$, we can take
\begin{align}
a=\frac{m}{\sqrt{2}} + \cO((mL)^2)\qquad\qquad\text{and}\qquad\qquad b=\frac{\sqrt{d}}{(d-1)^{3/2}}mL^{-1}\sqrt{d-1+(mL)^2}+\cO ( (mL)^2)~,
\end{align}
so $A_{0\mu}$ and $\bar{\pi}_0$ directly source $J_T^\mu$ and $\cO$, respectively. As $m\rightarrow 0$, $\tilh_{0\mu\nu}$ decouples from $J^T_\mu$ and $\cO$, and sources only the transverse traceless EMT. 
The coupling between  $A_{0\mu}$ and $\d_\mu\cO$ remains finite, however, in line with our expectation that $A_M$ and $\bar{\pi}$ together source a single, non-conserved current in the CFT.


In sum, we have derived a decomposition for the operator $\tau_{\mu\nu}$ which maps onto the conformal EMT and a decoupled vector current in the massless case. For small but finite mass, the operators $J_\mu$ and $T_{\mu\nu}$ mix, and the EMT is no longer conserved.

 \section{Two-point functions}
 
 We saw above that the CFT operator $\tau_{\mu\nu}$ is dual to a symmetric traceless tensor. In the limit of zero mass, we understood that $\tau_{\mu\nu}$ must contain a conserved traceless EMT, and a  non-conserved current $J_\mu$.
 Let us now study the simplest CFT data available to us, the two-point functions of $\tau_{\mu\nu}$
originally derived in \cite{Polishchuk:1999nh}:
\begin{align}\label{eq:tautauposition}
\langle \tau_{\mu\nu}({\bf x}) \tau_{\rho\sigma}({\bf y})   \rangle =\frac{C_{d,\tilde{\Delta}}}{|{\bf x}-{\bf y}|^{2\tilde{\Delta}+d}}\left[\frac{1}{2}J_{\mu\rho}({\bf x}-{\bf y})J_{\nu\sigma}({\bf x}-{\bf y})+\frac{1}{2}J_{\mu\sigma}({\bf x}-{\bf y})J_{\nu\rho}({\bf x}-{\bf y})-\frac{1}{d}\eta_{\mu\nu}\eta_{\rho\sigma} \right]~,
\end{align}
where
\begin{align}
J_{\mu\nu}({\bf x})=\eta_{\mu\nu}-2\frac{x_\mu x_\nu}{|{\bf x}|^2}\qquad \text{and}\qquad C_{d,\tilde{\Delta}}=\frac{\tilde{\Delta} (\tilde{\Delta}+\frac{d}{2}+1)\Gamma(\tilde{\Delta}+\frac{d}{2})}{2\pi^{d/2}\kappa^2(\tilde{\Delta}+\frac{d}{2}-1)\Gamma(\tilde{\Delta})} ~,
\end{align}
with
\begin{align}
\tilde{\Delta}=\sqrt{(mL)^2+\frac{d^2}{4}}~.
\end{align}
Henceforth we will absorb all factors of $M_P$ into the definition of the operators, which amounts to setting $M_P=1$. 

Note that we did not explicitly impose any tracelessness condition on  $h_{MN}$ (or its boundary values) -- nevertheless the two-point function is traceless. Technically speaking this is due to a suppression in $z$ of the trace part of $h_{\mu\nu}$, which in turn means that there is no coupling to the trace parts of the boundary sources. We have also seen, from the CFT side, that tracelessness of $\tau_{\mu\nu}$ is essential to matching DOF.


To analyze the two-point function in greater detail, we work in momentum space, where we can decompose \eqref{eq:tautauposition} in terms of traceless Lorentz structures as
\begin{align}\label{eq:tautaumomentum}
\langle \tau_{\mu\nu}(q)\tau_{\rho\sigma}(p)\rangle=\delta^d(p+q)\left[\Pi_\alpha\bbP^A_{\mu\nu\rho\sigma}+\Pi_\beta\bbP^B_{\mu\nu\rho\sigma}+\Pi_\gamma\bbP^C_{\mu\nu\rho\sigma}\right]~,
\end{align}
with form factors
\begin{align}\label{eq:polishform}
\Pi_\alpha&=\frac{Q(p)}{(\tilde{\Delta}+\frac{d}{2})(\tilde{\Delta}+\frac{d}{2}-1)}\left(\tilde{\Delta}^2-(d-1)\tilde{\Delta}+\frac{d}{2}\left( \frac{d}{2}-1\right) \right)~,\\
\Pi_\beta&=\frac{Q(p)}{(\tilde{\Delta}+\frac{d}{2})}\left(\frac{d}{2}-\tilde{\Delta} \right)~,\\
\Pi_\gamma&=Q(p)~,
\end{align}
and
\begin{align}
Q(p)=-\frac{\tilde{\Delta}\Gamma(1-\tilde{\Delta}) p^{2\tilde{\Delta}}}{2^{2\tilde{\Delta}+1}\Gamma(1+\tilde{\Delta})}~.
\end{align}
The traceless Lorentz structures
\begin{align}
\bbP^A_{\mu\nu\rho\sigma}&=\frac{1}{d(d-1)}\left(\eta_{\mu\nu}-d\frac{p_\mu p_\nu}{p^2}  \right)\left(\eta_{\rho\sigma}-d\frac{p_\rho p_\sigma}{p^2}  \right) ~, \\
\bbP^B_{\mu\nu\rho\sigma}&=-\frac{1}{2}\left[ \left(\eta_{\mu\rho}-\frac{p_\mu p_\rho}{p^2}  \right)\left(\eta_{\nu\sigma}-\frac{p_\nu p_\sigma}{p^2}  \right)
+\left(\eta_{\mu\sigma}-\frac{p_\mu p_\sigma}{p^2}  \right)\left(\eta_{\nu\rho}-\frac{p_\nu p_\rho}{p^2}  \right)\right]\nonumber\\
&\qquad\qquad\qquad\qquad+\frac{1}{2}\left(\eta_{\mu\rho}\eta_{\nu\sigma}+\eta_{\mu\sigma}\eta_{\nu\rho} \right)-\frac{p_\mu p_\nu p_\rho p_\sigma}{p^4}~,  \\
\bbP^C_{\mu\nu\rho\sigma}&=\frac{1}{2}\left(\eta_{\mu\rho}\eta_{\nu\sigma}+\eta_{\mu\sigma}\eta_{\nu\rho}\right)-\frac{1}{d}\eta_{\mu\nu}\eta_{\rho\sigma}-\bbP^A-\bbP^B  
\end{align}
form an orthonormal basis for Lorentz structures with four indices. They obey
$\bbP^\mu_{\phantom{\mu}\mu\rho\sigma}=\bbP^{\phantom{\mu\nu}\rho}_{\mu\nu\phantom{\rho}\rho}=0$, $\bbP_{\mu\nu\rho\sigma}=\bbP_{\nu\mu\rho\sigma}=\bbP_{\rho\sigma\mu\nu}$, where 
 $\bbP^I\bbP^J=\delta^{IJ}\bbP^I$. 
 
Recalling that in momentum space
\begin{align}
\tau_{\mu\nu}(p)=T^{Tt}_{\mu\nu}+a\frac{i}{p^2}p_{(\mu}J^T_{\nu)}-\frac{b}{2}\frac{1}{p^2}\left( \eta_{\mu\nu}p^2-dp_\mu p_\nu\right)\cO~,
\end{align}
we can extract $T^{Tt}_{\mu\nu}$ using $\bbP^C$:
\begin{align}
T^{Tt}_{\mu\nu}=\bbP_{\mu\nu}^{C\phantom{\nu}\rho\sigma}\tau_{\rho\sigma}~,
\end{align}
and $J^T_\mu,~\cO$ with
\begin{align}
\bbP_{\mu\nu}^{A\phantom{\nu}\rho\sigma}\tau_{\rho\sigma}=-\frac{b}{2}\frac{1}{p^2}\left( \eta_{\mu\nu}p^2-dp_\mu p_\nu\right)\cO~,
\end{align}
and
\begin{align}
\bbP_{\mu\nu}^{B\phantom{\nu}\rho\sigma}\tau_{\rho\sigma}=-a\frac{i}{p^2}p_{(\mu}J_{\nu)}^T~.
\end{align}

Applying these projectors $\bbP^I$ to the correlation function in momentum space, we recover all possible two-point functions among $T^{tT}$, $J^T$, and $\cO$. 
In particular, due to the orthogonality of the projectors, we can quickly see that there are no mixed two-point functions. For instance
\begin{align}
\langle T^T_{\mu\nu}(p)\left(iq_{(\rho}J^T_{\sigma)}(q) \right)\rangle = \delta^d (p+q)  \bbP^C\langle \tau\tau\rangle \bbP^B\propto\bbP^C\bbP^B=0~.
\end{align}
The two-point functions are simply
\begin{align}
\langle T^{Tt}_{\mu\nu}(p)T^{Tt}_{\rho\sigma}(q)\rangle&= \delta^{(d)}(p+q)\Pi_\gamma \bbP^C_{\mu\nu\rho\sigma} ~,\label{eq:TT}\\
\frac{1}{p^2}\langle p_{(\mu}J^T_{\nu)}(p)  q_{(\rho}J^T_{\sigma)}(q)\rangle &= -\delta^{(d)}(p+q)\frac{p^2}{a^2}\Pi_\beta \bbP^B_{\mu\nu\rho\sigma} = -\frac{2p^2}{m^2} \delta^{(d)}(p+q)\Pi_\beta \bbP^B_{\mu\nu\rho\sigma}~, \label{eq:JJ}\\
\label{eq:OO}\langle \cO(p)\cO (q) \rangle &= \delta^{(d)}(p+q)\frac{4}{d(d-1)b^2}\Pi_\alpha=\frac{4(d-1)^2}{d^2m^2L^{-2}(d-1+(mL)^2)} \delta^{(d)}(p+q) \Pi_\alpha~.
\end{align}
For small $m$, the leading terms in $\Pi_\beta$ and $\Pi_\alpha$ go like $(mL)^2$, so the $J^T_\mu$ and $\cO$ two-point functions given in \eqref{eq:JJ}, \eqref{eq:OO} are finite even in the
massless limit. The $T^{Tt}_{\mu\nu}$ two-point function is always order 1, and is not only traceless but transverse. 
When $m\ne 0$, $T^{Tt}_{\mu\nu}$ is still conserved in momentum space, but its conservation is
due entirely to the Lorentz structure $\bbP^C$, and is satisfied despite the fact that its two-point function scales as $p^{2\tilde{\Delta}}$ instead of the $p^{2d}$ of conserved, $d$-dimensional EMTs in local quantum field theories. 

We argue that this unusual scaling property, combined with conformality, implies that the dual of mGR is a nonlocal theory. Why exactly do conserved currents have a prescribed dimension? Consider, for instance, a $d$-dimensional system with non-abelian internal symmetry that has
corresponding vector currents $J^i_\mu$ ($i$ labels the generators $T^i$). The constant which parametrizes internal symmetry transformations is not dependent on spacetime at all, and is thus dimensionless. The 
 symmetry is generated by \textit{global} charges proportional to $T^i$. For local theories, these charges can be realized
as integrals over the time components of the conserved currents, and in fact we can write down a \textit{local} version of the global algebra:
\begin{equation}
[J^i_0(\vec{x}),J^j_0(\vec{y})]=if^{ijk}\delta^{(d-1)}(\vec{x}-\vec{y})J^k_0(\vec{x})~.
\end{equation}
In order for this relation to hold, the current must have dimension $(d-1)$.\footnote{This argument clearly applies only to non-abelian symmetry groups. Another argument is that in both the abelian and the non-abelian case, we can always choose to gauge the symmetry. The gauge field has mass dimension $1$ by definition, so the current to which it couples must have dimension $(d-1)$.} Similarly, the EMT generates spacetime translations, and is described by a parameter which has dimensions of length. 
By the identical argument to the above, then, we can see that for $d$-dimensional local theories, the EMT must have mass dimension $d$ if it is related to a conserved charge. In the dual to mGR, this is clearly not the case:
the fact that the scaling dimension of $T_{\mu\nu}$ differs from $d$ implies that there is \textit{no} relationship between the global charge and the current. So while the global conformal symmetry may remain, we 
cannot realize its commutation relations in terms of local currents. 

Furthermore,  the unusual scaling dimension of $T_{\mu\nu}$  cannot arise as the result of renormalization group (RG) flow from some local, Lorentz invariant UV fixed point. In that case, non-conserved operators would acquire anomalous dimensions, but the EMT would  retain the same scaling it possesses at $m=0$.

We argue in the next section that this strange behavior can arise in a conformal but non-local framework, sometimes termed an ``unparticle'' theory.

\section{Unparticles and toy models for anomalous scaling}

The ``unparticle stuff'' of \cite{Georgi:2007ek} (see also \cite{Krasnikov:1994fw,Krasnikov:2007fs}) defines a theory having a non-trivial IR fixed point, which couples weakly to the Standard Model (SM) at high energies.
Below some scale $\Lambda_U$, the new ``unparticle'' fields decouple from the SM, matching onto a generically strongly coupled conformal sector consisting of ``unparticle" operators.  Because this sector is conformally invariant, it cannot be characterized by excitations of well-defined mass. Instead, one can think of them as a non-integer number of states \cite{Georgi:2007ek}, or as fields
 with continuous mass distributions \cite{Krasnikov:1994fw,Krasnikov:2007fs}. The latter can arise as an effectively continuous KK tower,  which begs interpretation in terms of the AdS/CFT correspondence. Indeed, \cite{Stephanov:2007ry} demonstrated the emergence of  unparticle behavior in the CFT dual of a single massive scalar field on asymptotically AdS space. 
 
 We will now show that such theories, which effectively possess a fractional number of particles, are one way to produce the conserved, traceless tensors with non-integer scaling that we observed above.
 
Consider symmetric traceless two-tensor unparticles (``ungravitons'' in the language of \cite{Lee:2009dda})  $\cO_{\mu\nu}$. We can decompose the operator $\cO_{\mu\nu}$ in a basis of individual 
particle operators $\phi^{(i)}_{\mu\nu}$: $\cO_{\mu\nu}=\sum_{i=1}^n c_i\phi^{(i)}_{\mu\nu}$. Each $\phi^{(i)}$ produces a particle of mass $m_i$ and polarization $\epsilon_{\mu\nu}$ when it acts on the vacuum. 
We will eventually take $n\rightarrow \infty$: an continuum of particles with a continuous mass distribution.
The unparticle propagator has form
\begin{align}
\langle \cO_{\mu\nu}(p)\cO_{\rho\sigma}(0)\rangle = \sum\limits_{i=1}^\infty\frac{|c_i|^2}{p^2-m_i^2+i\epsilon}\left.\bbP^C_{\mu\nu\rho\sigma}\right|_{on-shell}=\int_0^\infty dt \frac{\rho_{\mu\nu\rho\sigma}(t)}{p^2-t+i\epsilon}~,
\end{align}
with spectral function $\rho_{\mu\nu\rho\sigma}=\sum_{i=1}^\infty |c_i|^2\delta(t-m_i)^2\bbP^C_{\mu\nu\rho\sigma}$. By ``on-shell'' we mean that the factors of momentum $p^2$ are replaced by $(-m_i^2)$ in the projector.

When the mass splitting vanishes, we achieve a spectral density $\rho(t)\sim t^{\Delta-1}$, which gives a propagator $\sim p^{2(\Delta-1)}$
for some (non-integer) $\Delta$ \cite{Krasnikov:2007fs}.  \cite{Stephanov:2007ry} details how such spectral functions can arise from scalar or vector fields in asymptotically AdS space. 
The interior of the $(d+1)$ dimensional space affects the spectrum and spectral densities of the CFT operators -- by tuning the background metric, then, it seems that one could produce a variety of different spectral functions. 

According to this picture, the bulk AdS theory corresponds to a conformal sector featuring unparticles generated by a vector current $J_\mu$ and a traceless, transverse two-tensor $T_{\mu\nu}^{tT}$. 
The theory is strongly coupled, so there is no straightforward way to capture the content of the unparticle sector, or in fact the relationship between $T^{tT}_{\mu\nu}$ and $J_\mu$. Nevertheless 
this theory is conformal: we know the \textit{dual} geometry still possesses the full AdS symmetry. It also seems to have dynamical gravity, which would imply the existence of an EMT -- but from the scaling of the correlation functions
 we understand that this cannot be the case. 

\subsection{Unparticle  actions}

Though the unparticle sector is strongly coupled by definition, we can write down a nonlocal action that generates the spectral functions defined above. This
very simple for scalar particles, because the Lorentz structure of the propagator is trivial:
\begin{align}
S_\chi=\frac{1}{2}\int d^dp ~ \chi(-p) p^{-2\delta}\chi(p)~,
\end{align}
where $\delta$ is some (usually non-integer) number. This generates a propagator of the form $\langle\chi\chi\rangle \sim ip^{2\delta}$. Similarly, we can reproduce the vector propagator by taking the action
\begin{align}
S_\zeta=\int d^dp \left\{ -\frac{1}{4}F^{(\zeta)}_{\mu\nu}p^{-2\delta} F^{(\zeta)\mu\nu} + p^{-2\delta}\zeta^\mu \cO_{gf\mu\nu}\zeta^\nu \right\}~,
\end{align}
where $F^{(\zeta)}_{\mu\nu}(p)= ip_\mu \zeta_\nu-ip_\nu \zeta_\mu$. The first term in the action still possesses the usual gauge freedom of Maxwell theory, so we include $\cO_{gf}(\zeta_\mu)$ to fix the gauge.
Say we choose 
\begin{align}
 \zeta^\mu \cO_{gf\mu\nu}\zeta^\nu=-\frac{1}{2\xi^2}(p^\mu \zeta_\mu)^2~.
\end{align}
The propagator for the vector unparticles is thus
\begin{align}
D^{(\zeta)}_{\mu\nu}=-ip^{2\delta-2}\left(\eta_{\mu\nu}-(1-\xi^2)\frac{p_\mu p_\nu}{p^2}\right)~.
\end{align}
Taking $\xi\rightarrow 0$ for example yields a transverse propagator.

Now consider spin 2 unparticles. We want the Lorentz structure of the propagator to be that of massive spin 2 particles, but for the scaling to differ from the usual $\Delta=d$. Consider the action
\begin{align}
S_\zeta=\int d^dp \ \theta^{\mu\nu}(-p)\left[\hat{\cE}_{\mu\nu\rho\sigma} +p^2 \mathcal{Q}_{gf}(\theta)\right]p^{-2\delta}\theta^{\rho\sigma}(p)~,
\end{align}
where $\hat{\cE}$ is the Lichnerowitz operator in momentum space (which is of order $p^2$) and $\mathcal{Q}_{gf}(\theta)$ again refers to some ``gauge fixing'' operator that eliminates the redundancy of the description. 
In order to match onto the propagator for the two-tensor unparticles, we should require
\begin{align}
p^{-2\delta}\left(\hat{\cE}+p^2\mathcal{Q}_{gf}(\theta) \right)^{\mu\nu\alpha\beta}\bbP^C_{\alpha\beta\rho\sigma}&=\frac{i}{2}\left( \delta^\mu_\rho\delta^\nu_\sigma+\delta^\mu_\sigma\delta^\nu_\rho\right)~,
\end{align}
where $\bbP^C_{\rho\sigma\mu\nu}$ is precisely the tensor structure appearing the two-point function, and
\begin{align}
\mathcal{Q}_{gf}=&-\frac{1}{4\gamma^2}\left[\frac{1}{p^2}(p_\mu p_\rho\eta_{\nu\sigma} +p_\mu p_\sigma\eta_{\nu\rho}+p_\nu p_\rho\eta_{\mu\sigma}+p_\nu p_\sigma\eta_{\mu\rho})+\beta\eta_{\mu\nu}\eta_{\rho\sigma} \right]~,
\end{align}
with $\gamma\rightarrow 0$ and $\beta$ an arbitrary non-zero constant yields the desired result.

We can now assemble these three different types of unparticles into a single  action, designed in such a way that the scaling dimensions of the ``un-propagators'' match those computed via holography: 
\begin{align}\label{eq:unparticle}
S_{total}&=\int d^dp \left\{K_\theta \theta^{\mu\nu}(-p)\left[\hat{\cE}_{\mu\nu\rho\sigma} +p^2 \mathcal{Q}_{gf}(\theta)\right]p^{-2(\tilde{\Delta}+1)}\theta^{\rho\sigma}(p)\right.\nonumber\\
&+\left. K_\zeta \left[ -\frac{1}{4}F^{(\zeta)}_{\mu\nu} p^{-2\tilde{\Delta}}F^{(\zeta)\mu\nu} + \zeta^\mu p^{-2\tilde{\Delta}-2}\cO_{gf\mu\nu}\zeta^\nu  \right] + K_\chi\chi(-p) p^{2(2-\tilde{\Delta})}\chi(p) \right\}~,
\end{align}
 where
 \begin{align}
 K_\theta^{-1}&=-\frac{\tilde{\Delta}\Gamma(1-\tilde{\Delta}) }{2^{2\tilde{\Delta}+1}\Gamma(1+\tilde{\Delta})}~,\\
 K_\zeta^{-1}&=-\frac{L^{-2}\tilde{\Delta}\Gamma(1-\tilde{\Delta}) }{2^{2\tilde{\Delta}}\Gamma(1+\tilde{\Delta})(\tilde{\Delta}+\frac{d}{2})^2}~,\\
 K_\chi^{-1}&=-\frac{L^4(d-1)^2}{d^2(\tilde{\Delta}+d-1)^2 \left(\tilde{\Delta}+\frac{d}{2}\right)}\frac{\tilde{\Delta}\Gamma(1-\tilde{\Delta}) }{2^{2\tilde{\Delta}-1}\Gamma(1+\tilde{\Delta})}~.
 \end{align}
 Note that this is a toy mock-up of a quadratic action that recovers the desired correlation functions we found using AdS/CFT.

\subsection{A nonlocal effective field theory from inhomogeneous scaling}
 
 In the previous subsection, we demonstrated a way to write down a $d$-dimensional action for unparticles to yield the Lorentz structures and scalings of the correlators derived from mGR. 
These theories preserve Lorentz (and in fact conformal) invariance, but are nonlocal -- in addition to being somewhat ad hoc. 
It is illuminating, then, to understand how such effective theories could arise from more local flat space theories. We will consider, in particular, $(d+1)$-dimensional flat space theories, which are local in $d$ dimensions (labeled by $x^\mu$), but which exhibit inhomogeneous (``Lifshitz-like'') scaling along one direction, $y$. These Lagrangians are still
invariant under (a modified set of ) diffeomorphisms. We eventually integrate out of the auxiliary ($y$) direction to find nonlocal conformally invariant field theories.

\subsubsection{Scalar particles}

Consider an action  for a scalar in half of $(d+1)$-dimensional Minkowski space. 
\begin{align}\label{eq:scalarLifshitz}
S=-\frac{1}{2}\int_0^\infty dy \int d^dx ~ \left[ \d_y^{\alpha_\Phi}\Phi(x,y)\d_y^{\alpha_\Phi}\Phi(x,y)+\d_\mu\Phi(x,y)\d^\mu\Phi (x,y) \right]+S_{bdy}~,
\end{align}
where the ``$(d+1)$''th coordinate $y\in[0,\infty)$.
Given appropriate boundary conditions for the fields and choice of $S_{bdy}$, this action has a well-defined variational principle. The parameter $\alpha_\Phi$ characterizes a theory inhomogeneous scaling in the auxiliary $y$ direction.
We assume that $\alpha_\Phi> 0$. The 
equations of motion are
\begin{align}
\delta S=\int_0^\infty dy \int d^dx ~& \left[ -(-1)^{-\alpha_\Phi}\delta\Phi(x,y)\d_y^{2\alpha_\Phi}\Phi(x,y)+\delta\Phi(x,y)\d^2\Phi (x,y) \right] \nonumber\\
&~~~~~~~-\int d^dx\sum\limits_{k=0}^{n-1} (-1)^{n-1+k-\alpha_\Phi}\left.\frac{d^{n-k-1}}{dy^{n-k-1}}\delta\Phi\frac{d^{k+2\alpha_\Phi-n}}{dy^{k+2\alpha_\Phi-n}}\Phi\right|_0^\infty~,
\end{align}
where $n$ is an integer such that  $n-1<\alpha_\Phi<n$.
We impose the boundary conditions that (1) $\Phi$ and all of its derivatives vanish as $y\rightarrow\infty$, and (2)  $ \delta\Phi(y=0)=0$, with $\Phi(x,0)=\varphi(x)$. 
We can eliminate any remaining boundary terms  at $y=0$ by adding a boundary action (\`{a} la Gibbons-Hawking-York):
\begin{align}
S_{bdy}=\frac{1}{2}\int dy\int d^dx ~ \d_y\left(\sum\limits_{k=0}^{n-2}(-1)^k \d_y^{n-k-1+\alpha_\Phi}\Phi\d_y^{k-n+\alpha_\Phi}\Phi \right)~.
\end{align}
Assuming (for technical reasons) that $\alpha_\Phi$ is a ratio of odd integers, we can show that
\begin{align}
\Phi(x,y)=e^{-y(-\d^2)^{1/2\alpha_\Phi}}\varphi(x)
\end{align}
solves the equations of motion in the $y$ direction. Plugging this solution into the action and integrating out $y$, we end up with the dynamics of the effectively $d$-dimensional field $\varphi(x)$:
\begin{align}
S_{on-shell,\Phi}&=-\frac{1}{2}\int d^dx \left.\Phi\d_y^{2\alpha_\Phi-1}\Phi\right|_0^\infty\nonumber\\
&=\frac{(-1)^{2\alpha_\Phi-2}}{2}\int d^dx ~ \varphi(-\d^2)^{1-\frac{1}{2\alpha_\Phi}}\varphi~.
\end{align}
In order to match the scaling of the scalar propagator derived in previous sections, we should choose $\alpha_\Phi=(2\tilde{\Delta}+2)^{-1}$.
Such scaling should of course be neutralized by a dimensionful parameter to achieve the correct scaling in
the Lagrangian.

\subsubsection{Vectors}
Similarly, we can construct a $(d+1)$-dimensional theory with inhomogeneous scaling for vectors, that leads to a non-local but conformal-invariant effective action upon dimensional reduction. The vector case is a bit more complicated, because
it is not sufficient to achieve the correct scaling, we also want to extract $d$-dimensional modes which are transverse. To this end,
let us introduce the gauge transformations
\begin{align}
\mathcal{A}_\mu\rightarrow \mathcal{A}_\mu + \d_\mu\mathcal{G}~,\\
\mathcal{A}_y\rightarrow \mathcal{A}_y + \d_y^{\alpha_A} \mathcal{G}~.
\end{align}
Under these transformations, the modified field strength $\tF_{MN}$ is invariant:
\begin{align}
\tF_{\mu\nu}&=\d_\mu \mathcal{A}_\nu-\d_\nu \mathcal{A}_\mu~,\\
\tF_{\mu y}&=\d_\mu \mathcal{A}_y - \d_y^{\alpha_A} \mathcal{A}_\mu~,
\end{align}
and we can write the action
\begin{align}
S=\int d^dx \int_0^\infty dy \left[ -\frac{1}{4}\tF_{MN}\tF^{MN}\right] + S_{bdy}
\end{align}
where $S_{bdy}$ denotes the boundary action determined as before to cancel boundary variations, given boundary conditions $\delta \mathcal{A}_\mu(y=0)=0$ and $\mathcal{A}_\mu\rightarrow 0$ as $y\rightarrow\infty$. 

Using the gauge freedom, we set $\mathcal{A}_y=0$, with which the $\mathcal{A}_y$ EOM  yields $\d_y^{\alpha_A}\d_\mu \mathcal{A}^\mu=0$. The longitudinal part of $ \mathcal{A}_\mu$ thus has no dynamics in the $y$ direction, and we take it to vanish. The EOM now take on a similar form to the ones we found in the scalar case:
\begin{align}
\d^2 \mathcal{A}_\mu^\perp = (-1)^{-\alpha_A}\d_y ^{2\alpha_A} \mathcal{A}_\mu^\perp~,
\end{align}
with similar solutions.
The on-shell effective action in $d$ dimensions becomes
\begin{align}
S_{on-shell,A}=\int d^dx \frac{(-1)^{2\alpha_A-2}}{2}\mathcal{A}_\mu^\perp (\d^2)^{1-\frac{1}{2\alpha_A}} \mathcal{A}_\mu^\perp~.
\end{align}
Matching dimensions to the transverse part of the vector current two-point function calculated holographically, we have $\alpha_A=(2\tilde{\Delta}+4)^{-1}$.

\subsubsection{Spin 2}
For spin 2, we can use a similar strategy to the vector. We will write the action for the tensor $\mathcal{H}_{MN}$ in terms of modified lapse and shift functions
\begin{align}
N&=(g^{yy})^{-1/2}\approx 1+\frac{1}{2} \mathcal{H}_{yy}+\dots~,\\
N_\mu&=g_{y\mu}\approx \mathcal{H}_{y\mu}+\dots~,
\end{align}
where now
\begin{align}
S=\int d^dx \int_0^\infty dy ~ N\sqrt{g_d}\left( R_d+\tilde{K}_{\mu\nu}^2-\tilde{K}^2\right)~,
\end{align}
and $\tilde{K}_{\mu\nu}$ is a modified extrinsic curvature,  which to linearized order in $\mathcal{H}_{MN}$ is 
\begin{align}
\tilde{K}_{\mu\nu}\approx\frac{1}{2}\left( \d_y^{\alpha_h} \mathcal{H}_{\mu\nu}-\d_\mu \mathcal{H}_{\nu y}-\d_\nu \mathcal{H}_{\mu y}\right)~.
\end{align}
%
The action is symmetric under a modified set of diffeomorphisms,
\begin{align}
\mathcal{H}_{\mu\nu}\rightarrow& \mathcal{H}_{\mu\nu} + \d_\mu\xi_\nu+\d_\nu\xi_\mu~,\\
\mathcal{H}_{\mu y}\rightarrow & \mathcal{H}_{\mu y}+\d_y^{\alpha_h} \xi_\mu~.
\end{align}
$\mathcal{H}_{yy}$ does not transform, and has no dynamics in the $y$ direction.
We have a total of $d$ gauge parameters defining the transformation, we expect the number of dynamical degrees of freedom to be 
\begin{align}
\frac{(d+1)(d+2)}{2}-1-2d = \frac{1}{2}d(d-1)~.
\end{align}


It will be convenient to decompose the fields as
\begin{align}
\mathcal{H}_{\mu\nu}(x,y)&=\mathcal{H}_{\mu\nu}^{Tt}+\d_\mu \mathcal{A}_\nu+\d_\nu \mathcal{A}_\mu+\d_\mu\d_\nu\pi+\eta_{\mu\nu}\sigma~,\\
\mathcal{H}_{\mu y}(x,y)&=\mathcal{B}_\mu +\d_\mu\varphi~,
\end{align}
where $\mathcal{H}_{\mu\nu}^{Tt}$ is transverse traceless, and $\mathcal{A}_\mu$, $\mathcal{B}_\mu$ are transverse.
We can use the gauge freedom to set $\mathcal{A}_\mu=0$ and $\pi=0$. Under this gauge fixing and after some manipulation, the EOM in $d+1$ dimensions are
\begin{align}\label{eq:htt}
\d^2 \mathcal{H}_{\mu\nu}^{Tt}&-(-1)^{-\alpha_h}\d_y^{2\alpha} \mathcal{H}_{\mu\nu}^{Tt} =\nonumber\\
&= \d_\mu\d_\nu \left[ \mathcal{H}_{yy}-(d-2)\sigma+2(-1)^{-\alpha_h}\d_y^{\alpha_h}\varphi\right]+\eta_{\mu\nu}(-1)^{1-\alpha_h}(d-1)\d_y^{2\alpha}\sigma\nonumber\\
&\qquad\qquad\qquad\qquad\qquad\qquad+2(-1)^{1-\alpha_h}\eta_{\mu\nu}\d_y^{\alpha_h}\d^2\varphi-(-1)^{-\alpha_h}\d_y^{\alpha_h}\left( \d_\mu \mathcal{B}_\nu +\d_\nu \mathcal{B}_{\mu}\right)~.
\end{align}
As before, we seek solutions to these equations from which we can derive the effective $d$-dimensional dynamics. Let us set $\sigma=0$. This implies $\d^2 \mathcal{B}_\mu\propto\sigma=0$, so we can assume $\mathcal{B}_\mu=0$ as well.
Now
\begin{align}
\d^2 \mathcal{H}_{\mu\nu}^{Tt}-(-1)^{-\alpha_h}\d_y^{2\alpha_h} \mathcal{H}_{\mu\nu}^{Tt}=2(-1)^{1-\alpha_h}\d_y^{\alpha_h}\eta_{\mu\nu}\d^2\varphi~.
\end{align}
Since the trace must vanish, we also have $\d_y^{2\alpha_h}\d^2\varphi=0$ and 
\begin{align}
\d^2 \mathcal{H}_{\mu\nu}^{Tt}-(-1)^{-\alpha_h}\d_y^{2\alpha_h} \mathcal{H}_{\mu\nu}^{Tt}=0~.
\end{align}
Imposing the boundary condition $\mathcal{H}_{\mu\nu}^{Tt}(x,0)=\mathcal{H}_{\mu\nu}^{Tt}(x)$, we have the same effective action,
\begin{align}
S_{on-shell, \mathcal{H}} =\int d^dx (-1)^{-\alpha_h}\mathcal{H}_{\mu\nu}^{Tt}\left( -\d^2\right)^{1-\frac{1}{2\alpha_h}}\mathcal{H}_{\mu\nu}^{Tt}~.
\end{align}
Here $\alpha_h=(2\tilde{\Delta}+2)^{-1}$ as for the scalar case.

\subsubsection{Assembling the effective action}

The goal of this exercise was to show that we can reproduce the correlators of the operator $\tau_{\mu\nu}$, or equivalently, the correlation functions of the unparticle theory in the previous subsection.  Putting together the results just derived, we have a 1-PI effective action in $d$ dimensions derived from the $(d+1)$-dimensional theory with anomalous scaling along an auxiliary direction. 
The $(d+1)$-dimensional Lagrangian is given by
\begin{align}
S_{d+1}=\int d^{d+1}x ~ \frac{1}{2}\left\{- (\d_y^\frac{1}{2(\tilde{\Delta}+1)}\Phi)^2 - \d_\mu\Phi\d_\mu\Phi -\frac{1}{2}\tilde{F}_{MN}\tilde{F}^{MN}+ N\sqrt{g_d}\left( R_d+\tilde{K}_{\mu\nu}^2-\tilde{K}^2\right) \right\}~,
\label{NLact}
\end{align}
where the inhomogeneous field strength, lapse and shift, and extrinsic curvatures are defined as above. Note that the action in $d$ of the $(d+1)$ dimensions is local-- it only exhibits unusual scaling in the auxiliary direction.  Furthermore, though the action 
(\ref{NLact})  contains ``gravity," it does not feature a cosmological term -- the background is not curved. We now assume that these bulk terms interact with some operators on the ``brane'' at $y=0$. 
Recall the decomposition of $\tau_{\mu\nu}$ as
\begin{align}
\tau_{\mu\nu} = \left(\bbP_{\mu\nu}^{A\rho\sigma} +\bbP_{\mu\nu}^{B\rho\sigma}+\bbP_{\mu\nu}^{C\rho\sigma} \right)\tau_{\rho\sigma}~.
\end{align}
in terms of idempotent, mutually orthogonal operators $\bbP^I$. We assume that
these (irreducible) operators living on the brane couple to $\Phi$, $\mathcal{A}$, $\mathcal{H}$. The couplings of the source $\tau_{\mu\nu}$ to the fields goes like
\begin{align}
S_{int} = \int d^dx \left[ C_A \frac{1}{\d^2}\left( \d^2\eta_{\mu\nu}-d\d_\mu\d_\nu\right)\Phi \bbP^{\mu\nu}_{A\rho\sigma}\tau^{\rho\sigma} + C_B \frac{1}{\d^2}(\d_\mu A_\nu +\d_\nu A_\mu)\bbP^{\mu\nu}_{B\rho\sigma}\tau^{\rho\sigma}+C_Ch_{\mu\nu}\bbP^{\mu\nu}_{C\rho\sigma}\tau^{\rho\sigma}\right]~,
\end{align}
where the couplings $C_A$, $C_B$, and $C_C$ are determined via the coefficients appearing the two-point function as computed via AdS/CFT.  

Now, upon integrating out in $y$, we have the 1-PI effective action
\begin{align}
S_{eff} = (-1)^{-\frac{2\tilde{\Delta}+1}{\tilde{\Delta}+1}}\int d^dx \left[ \Phi (-\d^2)^{-\tilde{\Delta}}\Phi + (-1)^{-\frac{1}{(\tilde{\Delta}+1)(\tilde{\Delta}+2)}}A_\mu^\perp (-\d^2)^{-\tilde{\Delta}-1}  A^{\perp\mu} + h_{\mu\nu}^{Tt} (-\d^2)^{-\tilde{\Delta}}h^{Tt\mu\nu} \right]~.
\end{align}
We have thus shown how the nonlocal behavior in a field theory might arise. This 1-PI effective action is related to the unparticle description of  \eqref{eq:unparticle} in that it yields the same correlation functions for the irreducible parts of $\tau_{\mu\nu}$.
 
 \section{Conclusion and Outlook}
We have described the dual of massive gravity (with small $m$) on AdS with AdS reference metric, which we conjecture to be a nonlocal conformal theory characterized by a spin 2 operator. As $m\rightarrow 0$, this  operator  reduces to the conserved EMT  and a vector current.
We showed, furthermore, that it is possible to generate such theories as the dimensional reductions of frameworks with inhomogeneous scaling.

Our discussion is complementary to existing work on field theory duals for bigravity. Indeed, an appropriate limit of bigravity -- whose AdS/CFT dual was studied in the compelling work of \cite{Aharony:2006hz, Kiritsis:2006hy} -- might takes us to massive gravity alone. Bigravity theories feature two dynamical metrics, with two different Planck masses, each corresponding to the EMT of some field theory. The setup of \cite{Aharony:2006hz}, for instance, features two copies of AdS space -- each housing a massless graviton and a scalar field -- which share a boundary. A particular choice of boundary condition couples the two scalar fields, which in turn generate a mass for a linear combination of the two gravitons via loop corrections. The full theory now has one massive graviton ($g_{\mu\nu}$)  dual to some symmetric two-tensor operator, and a massless graviton ($f_{\mu\nu}$) dual to the conserved total EMT. One arrives at mGR by decoupling the metric $f_{\mu\nu}$ from the rest of the bigravity dynamics by taking its Planck mass to infinity, as shown in \cite{Apolo:2012gg}.  However, this still does not answer the question of what the dual theory of the decoupled system should look like, especially since mGR exists as a well-defined theory in its own right, not only as a limit of bigravity.

In this work, we studied the quadratic order Lagrangian. An important next step would be to understand the interpretation of mGR parameters in terms of CFT data, which would go hand in hand with an analysis of the  interaction terms. 
Top-down formulations of gauge-gravity duality have the advantage that we can precisely identify the relationship between parameters in AdS to  parameters in the CFT. For standard AdS$_5$/CFT$_4$, the large $M_P$, small curvature ($L^{-1}$) theory corresponds to a large $N_c$, large 't Hooft coupling $\lambda=g_{YM}^2N_c$ limit  of the CFT via the well-known relations (see e.g. \cite{Aharony:1999ti} for a review)
\begin{align}
\left(\frac{L}{l_s}\right)^4\sim\lambda\,,\qquad\qquad (M_{10}l_s)^8\sim \frac{N_c^2}{\lambda^{2}}=\frac{1}{g_{YM}^4}\qquad\quad\quad \Rightarrow\qquad\qquad LM_{10}\sim N_c^{1/4}~.
\end{align}
 $L$ is the AdS radius, $M_{10}$ is the 10d string theory Planck mass, and $l_s$ is the string length. Only relative scales are important, and the string length $l_s$ never appears on its own in the supergravity approximation (because all massive string excitations are infinitely massive). If we choose the 10d Planck mass as our reference scale, we can clearly see from these relations that the string scale $l_s^{-1}\sim g_{YM}^{1/2}M_{10}\gg M_{10}$ when the CFT  is strongly coupled  ($g_{YM}\gg 1$). Similarly, when $\sqrt{N_c}\gg 1$ the spacetime curvature of AdS is weak,
$R\sim L^{-2}\ll M_{10}^{-2}$. It is in this limit that we can use perturbative (super)gravity. 
 
We can, at most, treat mGR on AdS$_{d+1}$ as some effective theory which may (or may not) derive from a string theory, so any identification of mGR parameters with CFT parameters is thus conjectural at best, though as $m\rightarrow 0$ we should recover the pure gravity identifications. We might conjecture, then, that our low energy effective theory on the mGR side corresponds to a strongly coupled conformal theory, characterized by a coupling $\lambda$. Just as the AdS scale is dual to the magnitude of the CFT coupling, we conjecture that the ratio $m/M_P$ defines an \textit{additional} CFT coupling,  $\tilde{\lambda}$ which encodes the non-locality of the theory. Future work should explore the details of this relationship via higher-point correlation functions.

 mGR itself displays interesting phenomena, which beg CFT interpretation. For instance, here is generically a scale above which the \textit{gravitational} excitations in mGR become strongly coupled. This implies some parametric threshold in terms of the CFT parameters beyond which we cannot reliably use perturbative gravity. 
It seems naively that for small $m$, this regime is always out of reach.

In order to use a perturbative gravity description, one should  consider excitations with energies below some cutoff $\Lambda_{cut}(m,M_P,L)$. This is simply the lowest scale suppressing interactions that appear as we expand the potential $\mathcal{U}$. We can read off the scale from the Galileon-type interactions of the type
\begin{align}
\square \bar{\pi} (\d\bar{\pi})^2~,
\end{align}  
which, once we canonically normalize the fields, are suppressed by a coefficient
\begin{align}
\Lambda_{cut}^{-\frac{d+3}{2}} = M_P^{-\frac{d-1}{2}}\cdot\beta^2L^4\cdot\gamma \qquad\Rightarrow\qquad \Lambda_{cut}\sim M_P  \left[ \left(\frac{L^{-1}}{M_P}\right)^3\left(\frac{m}{M_P}\right)^{-1}\left( (mL)^2+d-1\right)^{3/2} \right]^{\frac{2}{d+3}}~.
\end{align}
If we take $d=3$, and $L\rightarrow\infty$ and $m$ finite, we recover the Minkowski space result \cite{deRham:2010ik}: $\Lambda_{cut}\sim (M_P m^2)^{1/3} $.

While the graviton mass is a small, i.e. $m\ll L^{-1}\ll M_P$ we are free to consider momenta of order the AdS scale without straying into the regime of strong gravitational interactions (for any dimension $d>1$):
\begin{align}
\frac{L^{-1}}{\Lambda_{cut}}\sim (mL)^{\frac{2}{d+3}}\left( \frac{L^{-1}}{M_P}\right)^{\frac{d-1}{d+3}} < 1~.
\end{align}
This means that the gravitational description of the CFT should be reliable, even for higher point functions of CFT operators.

If we consider a graviton mass near  the AdS scale $L^{-1}$, e.g. $m=\rho L^{-1}$ where $\rho\sim \cO(1)$,
\begin{align}
\frac{L^{-1}}{\Lambda_{cut}}\sim \left[ \rho^2\left( \rho^2+d-1 \right)^{-3}\right]^{\frac{1}{d+3}}\left( \frac{L^{-1}}{M_P} \right)^{\frac{d-1}{d+3}}~.
\end{align}
It is straightforward to check that this function is maximized at $\rho^2=(d-1)/2$. As long as $L^{-1}\ll M_P$ by a few orders of magnitude, this ratio remains less than unity (for dimensions 2, 3, 4). 
For values of $m$ up to the AdS scale, then, strong interactions in gravity are safely above $M_P$, and we can explore the full parameter space of the dual couplings $\lambda$ and $\tilde{\lambda}$. 
It would be interesting to explore these limits in greater detail, and to understand in particular what conspiracy of CFT parameters $\lambda$ and $\tilde{\lambda}$ induces strong coupling behavior in mGR.

In this work we considered the duality between mGR with small $m$, and conformal theories. One might also study  
other  mGR solutions which are only asymptotically AdS. 
Is there, for instance, a Hawking-Page transition in these
frameworks? And if so, how  is this reflected in the structure of the CFT?

\section*{Acknowledgements}
We are grateful to Mark Wyman for collaboration in the early stages of this work, and  
thank S. Gubser and J. Maldacena for useful comments.  SKD was partially supported by the NYU 
Postdoctoral and Transition Program for Academic Diversity Fellowship during the completion of this work, and is currently supported by NSF grant PHY-1316452.  GG  is  supported
by  NASA  grant  NNX12AF86G  S06 and   NSF  grant PHY-1316452.

\bibliographystyle{JHEP}
\bibliography{DGW}

\end{document}